\documentclass[aps,prl,final,twocolumn,letterpaper]{revtex4}

\usepackage{graphicx}   
\usepackage{import}                         
\usepackage{epstopdf}
\usepackage{amsmath} 
\usepackage{float} 
\usepackage{bm}
\usepackage{amssymb}
\usepackage{quotes}
\usepackage{indentfirst}
\usepackage{color}
\usepackage{transparent}
\usepackage{dcolumn}
\usepackage{braket}
\usepackage{multirow}
\usepackage{cancel} 
\usepackage{mdframed}
\usepackage{color}
\usepackage{bm}
\usepackage{dsfont}
\usepackage{slashed}
\usepackage{soul, color} 
\soulregister\ref{7}  
\soulregister\cite{7} 
\renewcommand{\st}[1]{}

\usepackage{xr}

\makeatletter
\newcommand*{\addFileDependency}[1]{
  \typeout{(#1)}
  \@addtofilelist{#1}
  \IfFileExists{#1}{}{\typeout{No file #1.}}
}
\makeatother

\usepackage{textcomp} 
\usepackage{xifthen}
\usepackage{xcolor}
\usepackage{etoolbox}
\newboolean{togglechanges} 

\setboolean{togglechanges}{false}

\newcommand{\comment}[1]{\ifbool{togglechanges}
    {#1}  
    {\textcolor{blue}{#1}}}

\usepackage{bibentry}

\usepackage{graphicx}
\usepackage{dcolumn}
\usepackage{bm}

\usepackage{textcomp} 

\begin{document}
\rmfamily

\title{Automated modal analysis of entanglement with bipartite self-configuring optics}

\author{Charles~Roques-Carmes$^{1,\S}$}
\email{chrc@stanford.edu}
\author{Aviv~Karnieli$^{1, \S}$}
\email{karnieli@stanford.edu}
\author{David~A.~B.~Miller$^{1}$}
\author{Shanhui~Fan$^{1}$}
\email{shanhui@stanford.edu}

\affiliation{$^{1}$ E. L. Ginzton Laboratory, Stanford University, 348 Via Pueblo, Stanford, CA 94305\looseness=-1
\\
$^{\S}$ denotes equal contribution.}



\clearpage 

\vspace*{-2em}



\begin{abstract}
Entanglement is a unique feature of quantum mechanics. In coupled systems of light and matter, entanglement manifests itself in the linear superposition of multipartite quantum states (e.g., parametrized by the multiple spatial, spectral, or temporal degrees of freedom of a light field). In bipartite systems, the Schmidt decomposition provides a modal decomposition of the entanglement structure over independent, separable states. Although ubiquitous as a mathematical tool to describe and measure entanglement, there exists no general efficient experimental method to decompose a bipartite quantum state onto its Schmidt modes. Here, we propose a method that relies on bipartite self-configuring optics that automatically ``learns'' the Schmidt decomposition of an arbitrary pure quantum state. Our method is agnostic to the degrees of freedom over which quantum entanglement is distributed and can reconstruct the Schmidt modes and values by variational optimization of the network's output powers or coincidences. We illustrate our method with numerical examples of spectral entanglement analysis for biphotons generated via spontaneous parametric down conversion and provide experimental guidelines for its realization, including the influence of losses and impurities. Our method provides a versatile and scalable way of analyzing entanglement in bipartite integrated quantum photonic systems.
\end{abstract}

\maketitle

Entanglement is a fundamental property of quantum mechanical systems~\cite{horodecki2009quantum}, underpinning quantum computing algorithms~\cite{shor1999polynomial}, cryptography~\cite{yin2020entanglement}, teleportation protocols~\cite{bouwmeester1997experimental}, and tests of non-local realism~\cite{aspect1982experimental, groblacher2007experimental}. Optics is a prime platform for the generation and manipulation of entanglement between the many degrees of freedom of light and matter systems~\cite{raimond2001manipulating, o2009photonic, wineland2013nobel, erhard2020advances}. A paradigmatic example of bipartite entangled states in optics are photon pairs generated via spontaneous parametric down conversion (SPDC), which have been the workhorse of quantum photonic logic, computing, and simulations~\cite{carolan2015universal, zhong2020quantum, bogdanov2004qutrit, zhang2024entanglement}. There has been a growing interest in controlling entanglement of photon pairs with engineered pumps and nonlinear crystals~\cite{burlakov1999polarization, fedorov2008spontaneous, fedorov2011entanglement, hurvitz2023frequency, shukhin2024two}. Such engineered quantum sources may be used to shape separable biphoton wavefunctions for heralding~\cite{dosseva2016shaping, meyer2017limits, graffitti2018design} and for quantum information processing~\cite{graffitti2020direct, morrison2022frequency,shukhin2024two}.

A successful implementation of quantum computing and simulation protocols entails an accurate description and characterization of entanglement. In pure bipartite quantum systems, measures of entanglement are provided by the Schmidt number and the von Neumann entropy~\cite{horodecki2009quantum}. Generally, the Schmidt decomposition of a pure bipartite state provides a convenient modal description of entanglement (as a linear superposition of separable and orthogonal ``Schmidt modes'')~\cite{fedorov2014schmidt, sharapova2015schmidt}. Performing the Schmidt decomposition becomes progressively challenging as the Hilbert space becomes larger and even continuous, as in the case of encoding quantum entanglement in the spatial and spectral correlations of photon pairs~\cite{just2013transverse, sharapova2015schmidt, graffitti2020direct, morrison2022frequency, shukhin2024two}. \color{black} 

The problem of efficient Schmidt decomposition is an active field of study~\cite{amooei2024efficient, bhattacharjee2022measurement, averchenko2020reconstructing, kulkarni2018angular, kulkarni2022classical}, important for applications such as high-dimensional quantum communication~\cite{amooei2024efficient,dixon2012quantum, luo2019quantum, sit2017high} and quantum imaging~\cite{asban2019quantum, fuenzalida2022resolution}. Assumptions on the spatial and spectral shape and symmetries of the driving pump field and nonlinear crystal in SPDC can approximate the Schmidt decomposition analytically~\cite{kulkarni2018angular, kulkarni2022classical, averchenko2020reconstructing,sharapova2015schmidt, sharapova2018bright, straupe2011angular, law2004analysis, jha2011partial, van2006effect, miatto2012spatial} and reduce the scaling of the number of required measurements~\cite{amooei2024efficient, bhattacharjee2022measurement}. However, to date there is no efficient method for measuring the Schmidt decomposition in the most general setting of high-dimensional entangled photon states. 


\color{black}

Programmable networks of Mach-Zehnder interferometers (MZIs)~\cite{bogaerts_programmable_2020, harris2018linear} are an ideal platform for integrated quantum photonics~\cite{o2009photonic}. They have enabled the realization of quantum key distribution~\cite{sibson2017chip}, variational quantum algorithms~\cite{peruzzo2014variational, carolan2020variational}, multidimensional and multiphoton entanglement processing~\cite{matthews2009manipulation, wang2018multidimensional}, and quantum walk simulators~\cite{harris2017quantum}. In these demonstrations, the enabling feature of programmable MZI networks is that they can impart arbitrary unitary operators on spatial modes of quantum optical states~\cite{miller2013self, miller2013self2, carolan2015universal}. 

Self-configuring programmable network architectures~\cite{miller2013self,miller2013self2,miller2019waves,seyedinnavadeh_determining_2023,MillerAnalyze2020,milanizadeh2021coherent,pai2023experimentally,miller2015perfect,miller2013establishing,roques2024measuring} additionally offer simple and progressive configuration of the MZIs based only on power minimizations or maximizations, adapting automatically to the problem of interest, sometimes even without calculations. They can learn modal representations of (quantum) optical fields (e.g., communication modes~\cite{miller2013establishing, miller2019waves, seyedinnavadeh_determining_2023} and natural modes of partially coherent light~\cite{roques2024measuring}). However, their potential in the modal analysis of entanglement in bipartite quantum optical systems has not been explored thus far.

Here, we propose a general method to automatically analyze entanglement in bipartite quantum optical systems using self-configuring optics. Our approach relies on bipartite self-configuring networks (BSCN) composed of MZI meshes that impart unitary transformations to the two subspaces of the bipartite system.
The average power or coincidence counts at the output of the BSCN are sequentially optimized, which results in the network having ``learned'' the Schmidt decomposition of an input state. Then, the Schmidt number can be directly measured at the output (as the number of ports with non-zero average power/coincidence), and the Schmidt modes of the system can be deduced directly from the network's settings. Our approach thus automatically discovers the Schmidt modes without presuming any prior knowledge about these modes. \textcolor{black}{Furthermore, our architecture sequentially learns the most important Schmidt modes first (corresponding to the largest Schmidt values). In the practically important case where only a few Schmidt modes dominate, our method offers a favorable linear scaling with the Hilbert space dimension, both in required physical hardware and number of measurements, compared to the quadratic scaling if one directly measures the entire density matrix in order to perform a Schmidt decomposition.}

We illustrate our method with numerical examples of photon pairs generated via SPDC in various phase-matching settings. We show that our method can automatically measure Schmidt modes of photon pairs while being robust to single photon losses. Since our method allows us to automatically learn a modal decomposition of entanglement in the system, we also discuss potential applications in modal shaping of separable light, generation of states with controlled entanglement, \textcolor{black}{and in entanglement distribution protocols for quantum communication through scattering media}. Our method paves the way to full characterization, processing, generation and \textcolor{black}{distribution} of entanglement in quantum optical states in a robust, scalable, and tunable manner. We envision that our approach could form a building block in integrated quantum optical architectures for computing and simulation where entanglement measurement and control plays an essential role.

\begin{figure}
\centering
\vspace{-0.2cm}
  \includegraphics[scale=0.65]{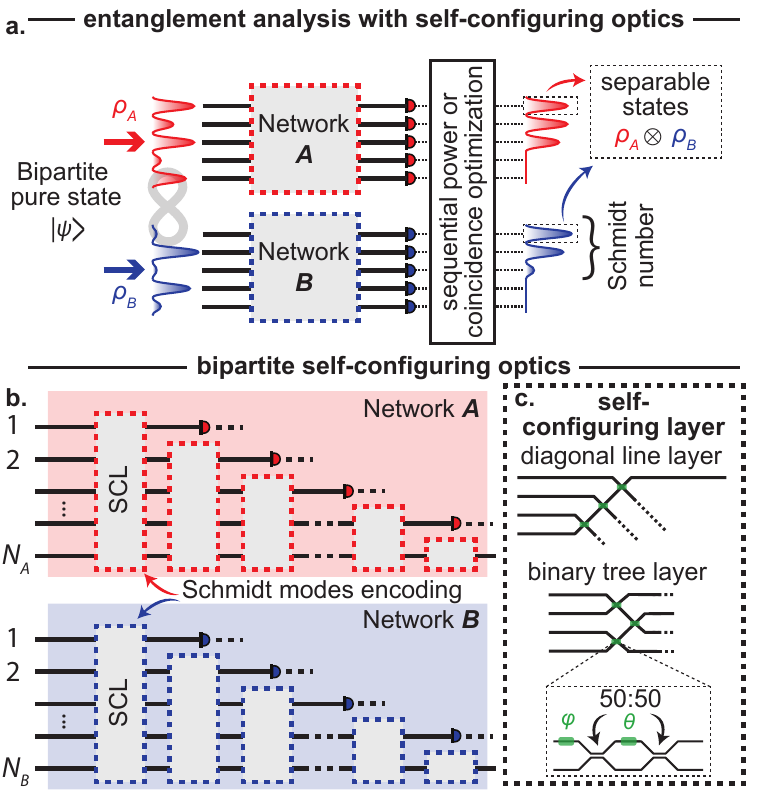}
    \caption{\small \textbf{Automatic Schmidt decomposition of bipartite quantum optical systems with self-configuring optics.} \textbf{a.} Conceptual schematic of automatic Schmidt decomposition enabled by bipartite self-configuring optics. A bipartite pure state $\ket{\psi_\text{in}}$ is input to two networks $A$ and $B$ acting upon Hilbert spaces $H_A$ and $H_B$, respectively. Their outputs (either power or coincidence counts) are sequentially optimized. The resulting output is a linear superposition of orthogonal separable states (Schmidt decomposition), and the Schmidt number can be read off as the number of non-zero outputs. \textbf{b.} Bipartite self-configuring network (BSCN) architecture. Each network consists of a cascade of self-configuring layers (SCLs). Each such SCL separates one mode to its output port (here, the top one in each layer), passing the remaining field to the next layer. \textbf{c.} SCL architecture, consisting of, e.g., a diagonal line or binary tree layer~\cite{MillerAnalyze2020}. Each node corresponds to a 2 by 2 Mach-Zehnder interferometer (MZI) parametrized by phases $(\varphi, \theta)$.}
    \label{fig:concept}
    \vspace{-0.3cm}
\end{figure}

\section*{Automatic Schmidt decomposition and entanglement analysis of bipartite quantum optical systems with self-configuring optics} 

We consider a pure state input $\ket{\psi_\text{in}}$ to the BSCN of Figure~\ref{fig:concept} (later, we shall consider photonic losses that might impair the state's purity). The input state is bipartite and is defined over the tensor product of two Hilbert spaces $H_A$ and $H_B$, with dimensions $N_A$ and $N_B$, respectively. These dimensions can be large, and even cover the (discretized) case of a continuous Hilbert space, e.g., as in spectral-bin encoding~\cite{morrison2022frequency, lu2023frequency}. A concrete example of this model consists of two single photons propagating over spatial modes of integrated networks $A$ (red) and $B$ (blue). We can decompose the state over an orthogonal basis $\ket{x_j}_A \ket{x_k}_B$ of $H_A \otimes H_B$ and write its Schmidt decomposition as:
\begin{align}
    \ket{\psi_\text{in}} &= \sum_{jk} G_{jk} \ket{x_j}_A \ket{x_k}_B \label{eq:psi-def}\\
    &= \sum_{i=1}^r \lambda_i \ket{y_i}_A \ket{y_i}_B, \label{eq:psi-schmidt}
\end{align}
where $G$ is a $N_A \times N_B$ state matrix whose singular value decomposition $G=UDV^\dagger$ maps to the input state's Schmidt decomposition~\cite{nielsen2010quantum} as $\ket{y_i}_A = \sum_j U_{ji} \ket{x_j}_A$, $\ket{y_i}_B = \sum_j V^*_{ki} \ket{x_k}_B$, and $D_{ii} = \lambda_i$ are the $r$ singular values of $G$ (ordered by decreasing value $\lambda_1 \geq \ldots \geq \lambda_r > 0)$. As can be seen from Equation~(\ref{eq:psi-schmidt}), the Schmidt decomposition provides a modal description of entanglement as a linear superposition of mutually orthogonal and separable states. The Schmidt number $r$, which quantifies the amount of entanglement in the state, corresponds to the number of non-zero $\lambda_i$'s. The Schmidt decomposition directly yields the von Neumann entropy of the input state: $S=-\sum_k \lambda_k^2 \log(\lambda_k^2)$.

The network shown in Figure~\ref{fig:concept} can automatically perform Schmidt decomposition of an input state $\ket{\psi_\text{in}}$ by simultaneous diagonalization of the reduced density operators of each subspace. The network is made of two cascades of self-configuring layers (SCLs) that form networks $A$ and $B$ (see Figure~\ref{fig:concept}(b)). SCL architectures can be defined topologically ~\cite{MillerAnalyze2020}; each SCL has a single output that is connected to each input by only one path through the MZI blocks~\footnote{Other unitary architectures could be used for these layers, but SCLs support simple configuration and have the minimum number of programmable elements.}.

These SCL cascades operate on spaces $H_A$ and $H_B$, respectively, transforming their reduced density matrices as $\rho^A \rightarrow U^A \rho^A (U^A)^\dagger$ and $\rho^B \rightarrow U^B \rho^B (U^B)^\dagger$ where $U^A$ and $U^B$ are reconfigurable unitary matrices imparted by networks $A$ and $B$, respectively. This SCL architecture allows for the implementation of sequential, layer-by-layer optimization methods~\cite{miller2013self,miller2013self2,roques2024measuring}. Example SCL architectures include diagonal lines or binary trees of MZIs~\cite{MillerAnalyze2020}, as shown in Figure~\ref{fig:concept}(c). These sequential, layer-by-layer optimizations over parameters of the BSCN can be realized by dithering~\cite{milanizadeh2021coherent,seyedinnavadeh_determining_2023} or \textit{in situ} back-propagation~\cite{hughes2018training, pai2023experimentally}.

We now show how the BSCN of Figure~\ref{fig:concept} can learn the Schmidt decomposition of the input state $\ket{\psi_\text{in}}$. Sequential optimization of the average output powers measured at port $k$ $\braket{P^A_k}$ ($k=1, \ldots, N_A$) and $\braket{P^B_k}$ ($k=1, \ldots, N_B$) maps to a variational definition of the eigendecomposition of $GG^\dagger$ and $(G^\dagger G)^T$ (see Supplementary Materials (SM), Sections~S1 and S2):
\begin{align}
    \underset{\tau^A_k}{\text{max}} \braket{P^A_k} &= \underset{u_k^A}{\text{max}}~(u_k^A)^\dagger ( G G^\dagger ) u_k^A = \lambda_k^2,\label{eq:pow-opt-A}\\
    \underset{\tau^B_k}{\text{max}} \braket{P^B_k} &= \underset{u_k^B}{\text{max}}~(u_k^B)^\dagger ( G ^\dagger G )^T u_k^B = \lambda_k^2,\label{eq:pow-opt-B}
\end{align}
where $\tau^A_k$ (resp., $\tau^B_k$) are the parameters of the $k$-th self-configuring layer of network $A$ (resp., $B$), and $u_k^A$ (resp. $u_k^B$) the $k$-th column of $(U^A)^\dagger$ (resp., $(U^B)^T$). One can see from Equations~(\ref{eq:pow-opt-A},~\ref{eq:pow-opt-B}) that tracing out subspace $B$ (resp., $A$) allows for the variational definition of the left (resp., right) singular vectors of $G$ and their respective singular values. 

Alternatively, sequential optimization of the single-photon coincidence counts between output ports $k$ of networks $A$ and $B$ (where $k = 1, \ldots, \text{min} (N_A, N_B)$) directly maps to a singular value decomposition of $G$ (see SM, Section~S2):
\begin{equation}
    \underset{\tau^A_k, \tau^B_k}{\text{max}} \braket{C_{kk}} = \underset{u_k^A, u_k^B}{\text{max}} |(u^A_k)^\dagger G u^B_k|^2 = \lambda_k^2,
\end{equation}
where the optimization runs over the $k$-th self-configuring layers of $A$ and $B$ simultaneously. Both optimization methods will result in setting the network parameters to that of the Schmidt modes, corresponding to the singular value decomposition of $G$, such that $U^A = U^\dagger$ and $U^B = V^T$.

Both sequential optimization methods provide an automatic way to find the Schmidt decomposition of the input state $\ket{\psi_\text{in}}$. The resulting settings of the BSCN allow one to ``read'' the Schmidt modes (corresponding, for instance, to the decomposition of the complex joint spectral amplitude (JSA) of a biphoton state), and the Schmidt number corresponds to half the number of output ports with non-zero average power, or the number of output port pairs of same index with non-zero coincidence. When considering two output ports sharing the same index $k$, $\ket{x_k}_A \ket{x_k}_B$, their joint output state is separable and corresponds to the \textcolor{black}{amplitude of the} Schmidt mode with singular value $\lambda_k$. \textcolor{black}{Individual Schmidt modes can be physically generated by feeding their corresponding output ports (while blocking the rest) as inputs to a pair of complementary networks applying $U_A^{\dagger}$ and $U_B^{\dagger}$ (as shown in Figure~\ref{fig:modes}).} A detailed derivation and description of the sequential optimization methods can be found in the SM, Sections~S1 and S2. 

\begin{figure}
\centering
\vspace{-0.2cm}
  \includegraphics[scale=0.75]{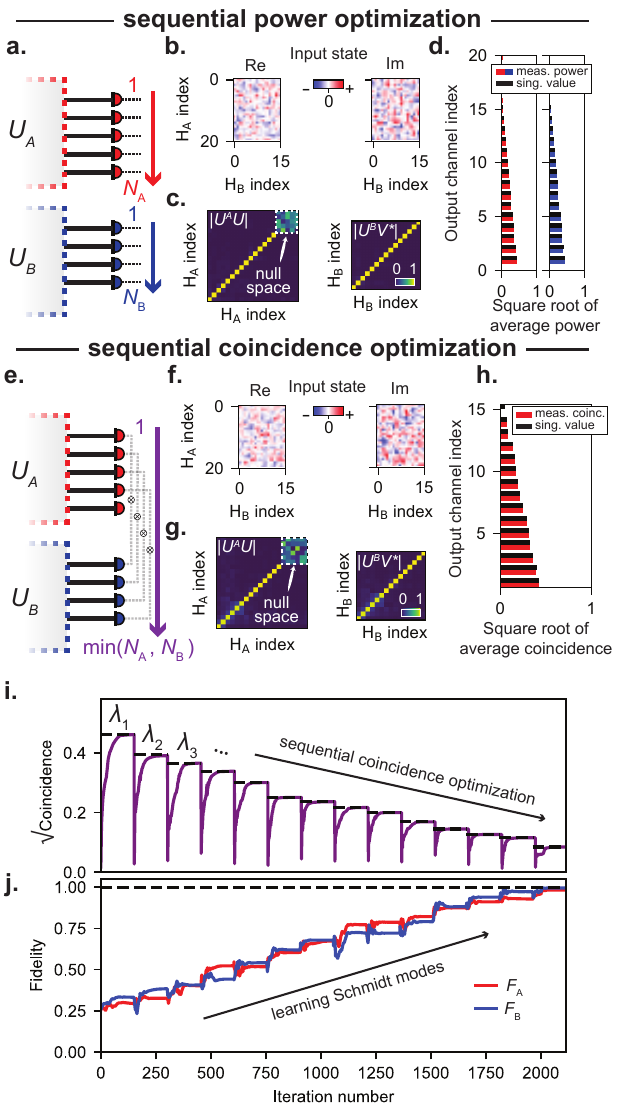}
     \caption{\small \textbf{Sequential optimization methods for entanglement analysis in self-configuring optics.} \textbf{a-d.} Sequential power optimization. \textbf{a.} Power optimization sequence: from $1$ to $N_A$ (on outputs of network $A$), then from $1$ to $N_B$ (on outputs of network $B$). \textbf{b.} Random matrix input example. \textbf{c.} Resulting eigenvector reconstruction from power optimization. \textbf{d.} Resulting simulated square roots of power outputs (meas. power) and singular values (sing. values) at the outputs of network $A$ (red) and $B$ (blue). \textbf{e-h.} Sequential coincidence counts optimization. \textbf{e.} Coincidence counts optimization sequence: from $1$ to $\text{min}(N_A, N_B)$ (on pairs of outputs ports from networks $A$ and $B$). \textbf{f.} Random matrix input example. \textbf{g.} Resulting eigenvector reconstruction from coincidence optimization. \textbf{h.} Resulting simulated square roots of coincidence counts (meas. coinc.) and singular values (sing. values). \textbf{i.} Square root of coincidence over iteration number, mapping to their corresponding Schmidt values. \textbf{j.} Fidelity of Schmidt modes in $A$ (red) and in $B$ (blue) over iteration number.}
    \label{fig:method}
    \vspace{-0.3cm}
\end{figure}

In the following, we turn to numerical examples illustrating our method, first with random matrices, to show the generality of this approach, and then the experimentally-relevant case of photon pairs generated via SPDC. 

In this first numerical example, we consider the performance of the two sequential optimization methods mentioned in the previous section. In both cases, gradients are calculated via automatic differentiation and optimization is performed via stochastic gradient descent~\cite{kingma2014adam}. The power optimization method (Figure~\ref{fig:method}(a)) runs over output ports $k$ of network $A$ for $k=1, \ldots, N_A$ and then $k$ of network $B$ for $k=1, \ldots, N_B$. We simulate the performance of the sequential power optimization over the outputs of the BSCN on a random matrix (shown in Figure~\ref{fig:method}(b)). First, the network learns the singular vectors of the state matrix $G$: this can be seen by plotting $U^A U$ and $U^B V^*$, which are equal to the identity matrix (except on the null space, see Figure~\ref{fig:method}(c)). The corresponding singular values can be read as the square root of the average power on the output ports of networks $A$ and $B$, as shown in Figure~\ref{fig:method}(d). The performance of the sequential coincidence maximization is shown on a similar random matrix $G$ in Figure~\ref{fig:method}(e-h), where similar performance is achieved, and the singular values can be measured as the square root of the average coincidence between ports $\ket{x_k}_A$ and $\ket{x_k}_B$. In these examples, the BSCN accurately identifies the Schmidt modes and values. 

\textcolor{black}{In our numerical simulations, optimization is performed with a modified version of stochastic gradient descent~\cite{kingma2014adam} and the performance of the algorithm in shown in Figure~\ref{fig:method}(i,j) for the sequential coincidence optimization. Each sequential coincidence (from $1$ to $\text{min}(N_A, N_B)$) quickly converges to the corresponding singular value. The fidelity increases over each sequential optimization step, and reaches values $\approx 1$ in our numerical experiments (for both Schmidt vectors in $A$ and $B$, see fidelity definition in the SM).}

\textcolor{black}{We now discuss the computational cost (both in terms of physical resources and number of measurements) for the BSCN to perform Schmidt decomposition (e.g., to find the Schmidt mode corresponding to the largest Schmidt value). Importantly, since both methods rely on Rayleigh quotient optimization, their convergence can be formally proven, as we show in the SM, Section~S7. This variational formulation of entanglement modal analysis also lends itself to stochastic gradient descent methods that have advantageous convergence speeds (which can be essentially independent of the state's dimensionality in some implementations~\cite{du2017gradient}). More generally, the number of measurements to achieve a given accuracy for a pair of Schmidt modes will scale like $O(N)$ (where $N$ is the Hilbert space dimension), and we show this linear scaling in simulations in the SI, Section~S7. Additionally, our method naturally orders the $r$ Schmidt values in decreasing order, such that only $O(rN)$ physical resources (number of MZIs) and $O(rN)$ measurements are in principle required, which can be a substantial improvement for high-dimensional systems with low Schmidt numbers, as compared to the worst-case $O(N^2)$ requirement using traditional Schmidt decomposition methods.}

\textcolor{black}{If single photons in either subspace are lost (e.g., absorbed or scattered), the input and output can mix with a vacuum state, which might result in errors in the power or coincidence measurements. In the SM, Section~S3 we model such detection and insertion losses as coupling each output and input port to a reservoir. We find that output losses do not affect the optimization process, and can be readily calibrated out by measuring the transmission matrix of the network (as is typically done in entanglement distribution experiments~\cite{lib2022quantum}). Non-uniform insertion losses at the input can distort the input state~\cite{miller2013establishing, roques2024measuring}. In that case, the coincidence optimization can still learn the correct SVD of the distorted input state traversing the network, while the sequential power optimization does not (nonetheless, in both cases there is enough information to reconstruct the original state, if the transmission matrix is known). This problem can be alleviated by equalizing insertion losses, preventing distortions, and allowing both the power and coincidence optimizations to learn the SVD of the input quantum state. }
Large losses may however result in reducing the number of counts on the detector and slow down the training process. Other MZI mesh architectures may also alleviate the influence of losses and imperfections~\cite{burgwal2017using, miller2015perfect}. This robustness to single photon losses provides further motivation to investigate the performance of our method in experimentally-relevant settings.

\begin{figure}
\centering
\vspace{-0.2cm}
  \includegraphics[scale=0.7]{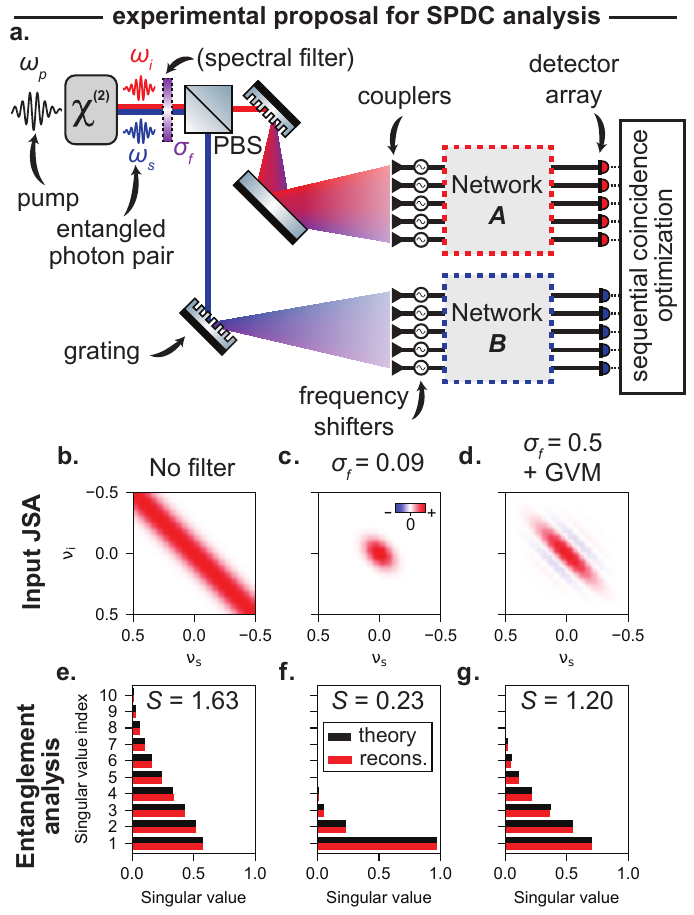}
    \caption{\small \textbf{Automatic entanglement analysis of spontaneous parametric down-converted (SPDC) photons.} \textbf{a.} Proposed experimental setup to automatically analyze entanglement of photon pairs generated via non-degenerate SPDC. A laser at frequency $\omega_p$ is pumping a second-order nonlinear crystal spontaneously generating entangled photon pairs at idler and signal frequencies $(\omega_i, \omega_s)$, separated by a polarization beam splitter (PBS). Different frequency components are routed to different input ports of two self-configuring networks (SCN) denoted $A$ ($\omega_i$) and $B$ ($\omega_s$). The SCN perform sequential coincidence counts maximization to automatically perform the Schmidt decomposition of the photon pair. \textbf{b-d.} Input joint spectral amplitude (JSA). \textbf{e-g.} Entanglement analysis with BSCN. $S$ denotes the von Neumann entropy. \textbf{b, e.} Numerical results for an unfiltered SPDC experiment ($\sigma_f \rightarrow \infty$, as defined in SM, Section~S4). \textbf{c, f.} Same as \textbf{b, e.}, but with a filter applied to the signal and idler modes ($\sigma_f = 0.09$). \textbf{d, g.} Same as \textbf{b, e.}, but with a filter ($\sigma_f = 0.2$) and group velocity mismatch (GVM) corrections. All frequencies are normalized to frequency range of interest: pump bandwidth is $\sigma = 0.1$ for (\textbf{b-c.}) and $\sigma = 0.5$ for (\textbf{d.}).}
    \label{fig:spdc}
    \vspace{-0.3cm}
\end{figure}

\begin{figure}
\centering
\vspace{-0.2cm}
  \includegraphics[scale=0.6]{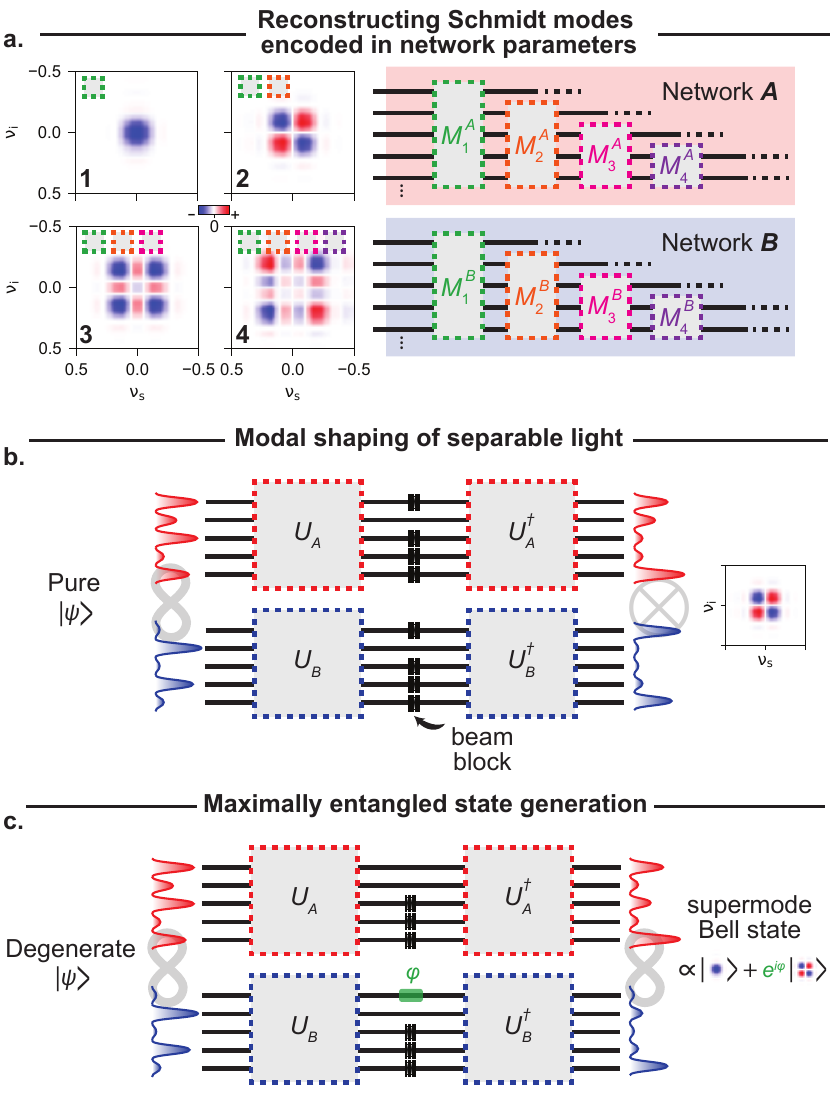}
    \caption{\small \textbf{Schmidt mode reconstruction and entangled light processing with bipartite self-configuring optics.} \textbf{a.} Left: First four Schmidt modes reconstructed from BSCN parameters in the SPDC example of Figure~\ref{fig:spdc}\textbf{d,g.}. Right: BSCN architecture for reference, where each self-configuring layer is highlighted in a different color. The self-configuring layers used to reconstruct the Schmidt modes on the left are represented by the number and color of gray squares as insets. \textbf{b.} Once trained, BSCN can be used to generate separable light with a given modal shape by blocking all output ports except the two ports corresponding to that mode index (in networks $A$ and $B$) \textcolor{black}{and feeding them as input to networks applying the inverse transformation}. \textbf{c.} If trained on a highly degenerate input state (such that two singular values are approximately equal), BSCN can also be used to generate supermode Bell states (by blocking all output ports except the four ports corresponding to degenerate singular values in each network). An additional phase shifter can be added on one of the output ports to control the relative phase between the two terms of the pair.}
    \label{fig:modes}
    \vspace{-0.3cm}
\end{figure}

\begin{figure}
\centering
\vspace{-0.2cm}
  \includegraphics[scale=0.7]{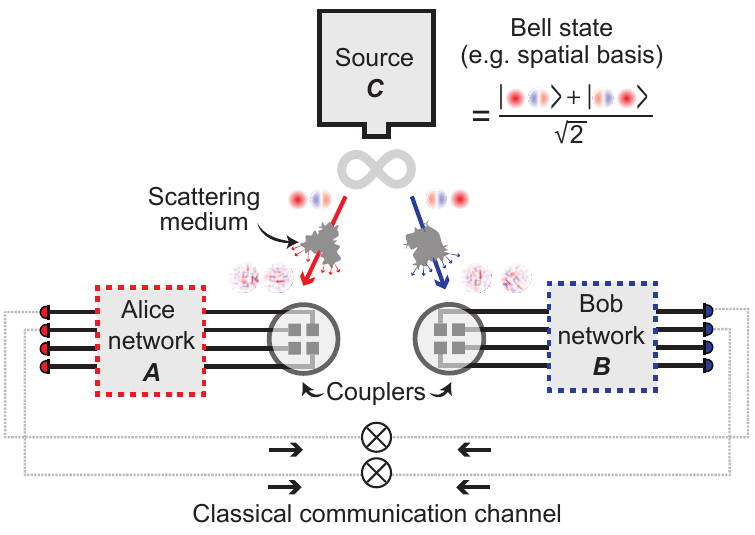}
    \caption{\textcolor{black}{\small \textbf{Entanglement distribution through scattering media with BSCN.} Proposed setup for quantum teleportation through scattering media with BSCN. A source $C$ is generating Bell states sent to Alice ($A$) and Bob ($B$) who analyze their single photon with their own SCN. A classical communication channel between the two allows them to perform coincidence measurements.}}
    \label{fig:teleportation}
    \vspace{-0.3cm}
\end{figure}

\section*{Analyzing and processing entanglement in spontaneously down-converted photon pairs} 

We now investigate the performance of our method in the analysis of entangled photon pairs generated via SPDC. \textcolor{black}{As noted in the previous section, our method is naturally suited for the analysis of spatial mode entanglement. To further generalize the applicability of our approach, we consider below the case of frequency bin entanglement. This concept could be readily generalized to other high-dimensional Hilbert spaces, e.g., correlations in spatial modes such as orbital angular momentum~\cite{mair2001entanglement, erhard2018twisted, kovlakov2018quantum}.} The proposed experimental setup that we model is shown in Figure~\ref{fig:spdc}(a). We consider a second-order nonlinear crystal pumped by a laser at frequency $\omega_p$ and spontaneously generating photon pairs at frequencies $\omega_s$ (signal) and $\omega_i$ (idler). The SPDC photons then go through an optional spectral filter to shape the joint spectral density (width $\sigma_f$) and a polarizing beam splitter to split signal and idler beams. \textcolor{black}{Gratings are then used to diffract spectral modes of the idler (signal) photons, occupying equidistant frequency bins centered around $\omega_{i(s),j} = \omega_{i(s),0} + j\Omega$, into corresponding spatial input modes $x_j$} (which couple to the network with a grating coupler). \textcolor{black}{The frequency bins, now occupying different spatial modes, are then shifted to a common frequency $\omega_{i(s),0}$ (see SM section S5). This can be done by an array of (on-chip) electro-optic frequency shifters~\cite{hu2021chip, zhu2022spectral, lingaraju2022bell, lu2023frequency} driven by corresponding $\Omega$ harmonics \footnote{Similarly, for analyzing orbital angular momentum entanglement, the gratings and frequency shifters can be replaced by an off-chip spatial mode sorter~\cite{fickler2014interface, lightman2017miniature} that converts orbital angular momentum modes into deflected Gaussian modes, which then couple to the chip.}. }  

In the absence of insertion and detection losses, and spectral mixing by the gratings and frequency shifters \textcolor{black}{(as shown in SM Section S3, and discussed above, the BSCN is robust to such imperfections)}, the resulting input state to the BSCN is:
\begin{equation}
    \ket{\psi_\text{in}} = \sum_{j,k} f(\nu_{s,j}, \nu_{i,k}) \ket{x_j}_A \ket{x_k}_B,
\end{equation}
where $f(\nu_{s}, \nu_{i})$ is the JSA of the photon pair at frequency $(\nu_{s}, \nu_{i})$, and $\nu_{s,j} = \omega_{s,j} - \omega_{s,0}$ and $\nu_{i,k} = \omega_{i,k} - \omega_{i,0}$ correspond to the discrete frequency bins around the central frequencies. The JSA can be controlled by shaping the incident pump spectrum~\cite{werner1995ultrashort, arzani2018versatile}, the nonlinear crystal poling pattern~\cite{graffitti2020direct, morrison2022frequency, shukhin2024two}, via dispersion engineering~\cite{u2006generation, CHRIST2013351}, or adiabatic frequency conversion~\cite{presutti2024highly}. Below, we shall focus on the ubiquitous scenarios of spectral filtering and group velocity mismatch between the pump, signal, and idler (Figure~\ref{fig:spdc}(d,g)). Further details on our experimental model can be found in SM, Section~S4. 

Several phase-matching settings are shown in Figures~\ref{fig:spdc}(b-d), as well as the BSCN output after convergence in Figures~\ref{fig:spdc}(e-g).
In all cases, the BSCN learns the Schmidt decomposition of the input state, and we can use the average coincidence counts to calculate the input state's von Neumann entropy (with negligible error compared to ground truth value $ < 1 \%$).
The addition of a spectral filter in Figure~\ref{fig:spdc}(c,f) reduces the entanglement by removing spectral components of the photon pair, namely making the state ``more separable.'' In general, group velocity mismatch is present (Figure~\ref{fig:spdc}(d,g)), resulting in anti-correlated  side lobes in the JSA. Several methods have been proposed to further shape the joint spectral density with engineered crystals~\cite{hurvitz2023frequency} and pumps~\cite{boucher2021engineering} and the BSCN could be also applied to perform modal analysis of entanglement in these cases as well. We note that the experimental proposal from Figure~\ref{fig:spdc} might be generalized to a fully integrated setup. This could be achieved by leveraging on-chip biphoton sources~\cite{luo2017chip, chapman2024chip} and would avoid coupling losses due to grating couplers.

A distinctive feature of our method is that the Schmidt modes (including their phases, \textcolor{black}{see SM, Section S2C}) are directly encoded into the parameters of the network after it has converged. Specifically, the $k$-th Schmidt mode is encoded in the network parameters (e.g., the phase shifts for various waveguide segments of MZIs) of layers $1, \ldots, k$ of both networks $A$ and $B$. We can therefore reconstruct the Schmidt modes (of the SPDC example from Figure~\ref{fig:spdc}(d)) as shown in Figure~\ref{fig:modes}(a). The self-configuring layers used for the reconstruction of each mode are indicated with dashed square boxes in each panel.

By automatically performing modal decomposition of the entanglement structure of the input state, the BSCN offers other possibilities in modal shaping and generation of entanglement states. Namely, by blocking all output ports except the ones corresponding to a given biphoton mode, \textcolor{black}{and using them as inputs for a pair of inverse networks (or even by reflecting them back through the output ports of the original networks)}, one can generate separable states with known shape (e.g., in spectral or spatial domain), as shown in Figure~\ref{fig:modes}(b). Additionally, if one inputs a degenerate quantum state (such that $\lambda_1\approx \lambda_2$, as can be done with broadband phase-matching~\cite{zielnicki2018joint} or engineered nonlinear crystals~\cite{hurvitz2023frequency}), a ``supermode Bell state'' can be generated by considering the output of the first two ports of both networks: $\ket{\psi_\text{out}} \propto \lambda_1 (\ket{x_1}_A \ket{x_1}_B + e^{i\varphi} \ket{x_2}_A \ket{x_2}_B)$, where an additional phase $\varphi$ can be imparted by a phase shifter on one of the output ports. 

\color{black}

\section*{Entanglement distribution through scattering media with BSCN}

By shaping the modal composition of entangled states, our approach can also be applied to quantum communication protocols. One famous example is that of quantum teleportation~\cite{bouwmeester1997experimental} between two agents Alice ($A$) and Bob ($B$), where a Bell state is provided by a third-party source $C$ (see Figure~\ref{fig:teleportation}). A successful implementation of quantum teleportation entails the distribution of entanglement between the two parties $A$ and $B$, or, in other words, communication of each photon of the Bell state to each of the two parties. In real-life settings, this communication may experience scattering and mode mixing through different communication channels (e.g., due to random scattering). The scattering of quantum light through complex media is an active field of research~\cite{lib2022quantum}, with immediate applications to quantum communication and entanglement distribution protocols through scattering channels such as turbulent atmosphere and multimode fibers. Conventional schemes compensate for the scattering through adaptive wavefront shaping of the photon pairs by characterizing the channel using an auxiliary laser at or near the same photon frequency~\cite{defienne2016two, defienne2018adaptive, wolterink2016programmable, cao2020long} or even the pump laser used to generate the photons~\cite{lib2020real}. 

The BSCN offers an alternative approach based on automatic Schmidt decomposition via sequential coincidence optimization. Specifically, if both Alice and Bob possess a SCN that they use to process the photon they receive, and perform the Schmidt learning method described in this work, they can learn the basis in which Alice should perform measurement to teleport her state to Bob. To do so, a classical communication channel is necessary between Alice and Bob, but there is no need to perform any characterizations of the communication channels  between the source to either Alice or Bob. More generally, the BSCN allows us to undo the influence of random scattering when distributing multimode entangled states. A formal derivation of this communication protocol is provided in the SM, Section~S8.

In fact, in this protocol the BSCN finds the effective communication modes~\cite{miller2013establishing, seyedinnavadeh_determining_2023} between Alice and Bob. This idea can therefore be generalized to the Klyshko configuration~\cite{shekel2024shaping}, which replaces one of the photon detectors (say, Alice’s) by a classical input laser beam to characterize the effective scattering channel between Alice, the source, and Bob. Characterizing the communication channel in this manner to configure the BSCN could boost the convergence speed and detection bandwidth of the protocol, which can be critical for entanglement distribution through rapidly fluctuating media.

\color{black}
\section*{Discussion}

We further discuss potential applications of our method and experimental considerations for their realization. 

Our methods could be applied to entanglement analysis in other quantum systems, such as temporal entanglement in atom-photon coupling
\cite{bogdanov2006schmidt} and macroscopic quantum states of light, such as bright squeezed vacuum~\cite{sharapova2015schmidt, sharapova2018bright}. In both cases, multimodal entanglement over temporal, spatial, or spectral degrees of freedom can be described on the Schmidt basis which reveals the fundamental mechanisms of entanglement generation. Many of these systems, including photon pairs from SPDC, rely on several control parameters of light or matter, which can be dynamically optimized to shape or maximize entanglement. Since the BSCN automatically learns the Schmidt decomposition of the input state, it could also serve as a form of feedback on the system's control parameters to learn their influence on the entanglement structure of the input state. In particular, well established techniques for shaping photon pairs~\cite{u2006generation, dosseva2016shaping, meyer2017limits, graffitti2020direct, morrison2022frequency, hurvitz2023frequency, shukhin2024two} are often sensitive to experimental parameters such as beam alignment, poling fabrication, and even temperature. Our proposal could thus allow for robustness and tunability even in the presence of experimental imperfections.

The automatic learning of the input state's Schmidt decomposition requires a pure input state of the network $\ket{\psi_\text{in}}$. Several sources of noise and imperfections may result in impurity of the quantum state. First, stochastic losses through the network yield a mixture of pure states with different photon numbers. With the above-mentioned loss model, we show that post-selection on a given number state (e.g., by performing coincidence or single photon counts) mitigates the influence of this impurity on the performance of our algorithm (see SM, Section~S3). Second, the input state might be mixed with other (random) states, resulting in a statistical mixture. In general, this would result in the emergence of cross-coincidence counts between output ports $\ket{x_j}_A$ and $\ket{x_k}_B$ where $j\neq k$ after diagonalization of the reduced density matrices performed by the BSCN (see SM, Section~S6). Practically, this means that the BSCN might also be used to ``detect'' impurities in the input quantum state in this more general setting. 

\color{black}

Our proposal provides a high-dimensional (two-qudit) photonic-hardware implementation of variational quantum singular value decomposition circuits~\cite{bravo2020quantum, wang2021variational} which decompose bipartite entanglement of general multiple-qubit states. As a future direction, it will be interesting to generalize our architecture to bipartite entangled states of more than two photons, allowing for multiple-qudit variational quantum singular value decomposition circuits.

\color{black}

In conclusion, we have presented a general and versatile method to automatically analyze the entanglement of a multimode quantum optical state. Our method relies on a variational definition of the Schmidt decomposition performed by a BSCN. The methods presented in this work could be applied to a wide range of quantum optical systems where the engineering and shaping of entanglement plays an essential role.


\section{Ethics approval and consent to participate}
Not applicable.

\section{Availability of data and materials}
The data and codes that support the plots within this paper and other findings of this study are available from the corresponding authors upon reasonable request. Correspondence and requests for materials should be addressed to C.~R.-C. (chrc@stanford.edu) and A.~K. (karnieli@stanford.edu).

\section{Competing interests}
The authors declare no potential competing financial interests.

\section{Funding}

C.~R.-C. is supported by a Stanford Science Fellowship. A.K. is supported by the VATAT-Quantum fellowship by the Israel Council for Higher Education; the Urbanek-Chodorow postdoctoral fellowship by the Department of Applied Physics at Stanford University; the Zuckerman STEM leadership postdoctoral program; and the Viterbi fellowship by the Technion. S.~F. and D.A.B.M. acknowledge support by the Air Force Office of Scientific Research (AFOSR, grant FA9550-21-1-0312). D.A.B.M. also acknowledges support by the Air Force Office of Scientific Research (AFOSR, grant FA9550-23-1-0307).

\section{Authors' contributions}

C.R.-C., A.K., D.A.B.M., and S.F. conceived the idea. C.R.-C. and A.K. performed the numerical experiments. C.R.-C. and A.K. wrote the manuscript with inputs from all authors. S.F. and D.A.B.M. supervised the research.

\section{Acknowledgements}
The authors would like to thank Paul-Alexis Mor, Carson Valdez, Annie Kroo, Anna Miller, and Olav Solgaard for stimulating conversations. 

\bibliographystyle{ieeetr}
\bibliography{bibliography}

\begin{thebibliography}{10}

\bibitem{horodecki2009quantum}
R.~Horodecki, P.~Horodecki, M.~Horodecki, and K.~Horodecki, ``Quantum entanglement,'' {\em Reviews of modern physics}, vol.~81, no.~2, p.~865, 2009.

\bibitem{shor1999polynomial}
P.~W. Shor, ``Polynomial-time algorithms for prime factorization and discrete logarithms on a quantum computer,'' {\em SIAM review}, vol.~41, no.~2, pp.~303--332, 1999.

\bibitem{yin2020entanglement}
J.~Yin, Y.-H. Li, S.-K. Liao, M.~Yang, Y.~Cao, L.~Zhang, J.-G. Ren, W.-Q. Cai, W.-Y. Liu, S.-L. Li, {\em et~al.}, ``Entanglement-based secure quantum cryptography over 1,120 kilometres,'' {\em Nature}, vol.~582, no.~7813, pp.~501--505, 2020.

\bibitem{bouwmeester1997experimental}
D.~Bouwmeester, J.-W. Pan, K.~Mattle, M.~Eibl, H.~Weinfurter, and A.~Zeilinger, ``Experimental quantum teleportation,'' {\em Nature}, vol.~390, no.~6660, pp.~575--579, 1997.

\bibitem{aspect1982experimental}
A.~Aspect, J.~Dalibard, and G.~Roger, ``Experimental test of bell's inequalities using time-varying analyzers,'' {\em Physical review letters}, vol.~49, no.~25, p.~1804, 1982.

\bibitem{groblacher2007experimental}
S.~Gr{\"o}blacher, T.~Paterek, R.~Kaltenbaek, {\v{C}}.~Brukner, M.~{\.Z}ukowski, M.~Aspelmeyer, and A.~Zeilinger, ``An experimental test of non-local realism,'' {\em Nature}, vol.~446, no.~7138, pp.~871--875, 2007.

\bibitem{raimond2001manipulating}
J.-M. Raimond, M.~Brune, and S.~Haroche, ``Manipulating quantum entanglement with atoms and photons in a cavity,'' {\em Reviews of Modern Physics}, vol.~73, no.~3, p.~565, 2001.

\bibitem{o2009photonic}
J.~L. O'brien, A.~Furusawa, and J.~Vu{\v{c}}kovi{\'c}, ``Photonic quantum technologies,'' {\em Nature photonics}, vol.~3, no.~12, pp.~687--695, 2009.

\bibitem{wineland2013nobel}
D.~J. Wineland, ``Nobel lecture: Superposition, entanglement, and raising {S}chr{\"o}dinger’s cat,'' {\em Reviews of Modern Physics}, vol.~85, no.~3, p.~1103, 2013.

\bibitem{erhard2020advances}
M.~Erhard, M.~Krenn, and A.~Zeilinger, ``Advances in high-dimensional quantum entanglement,'' {\em Nature Reviews Physics}, vol.~2, no.~7, pp.~365--381, 2020.

\bibitem{carolan2015universal}
J.~Carolan, C.~Harrold, C.~Sparrow, E.~Mart{\'\i}n-L{\'o}pez, N.~J. Russell, J.~W. Silverstone, P.~J. Shadbolt, N.~Matsuda, M.~Oguma, M.~Itoh, {\em et~al.}, ``Universal linear optics,'' {\em Science}, vol.~349, no.~6249, pp.~711--716, 2015.

\bibitem{zhong2020quantum}
H.-S. Zhong, H.~Wang, Y.-H. Deng, M.-C. Chen, L.-C. Peng, Y.-H. Luo, J.~Qin, D.~Wu, X.~Ding, Y.~Hu, {\em et~al.}, ``Quantum computational advantage using photons,'' {\em Science}, vol.~370, no.~6523, pp.~1460--1463, 2020.

\bibitem{bogdanov2004qutrit}
Y.~I. Bogdanov, M.~Chekhova, S.~Kulik, G.~Maslennikov, A.~Zhukov, C.~Oh, and M.~Tey, ``Qutrit state engineering with biphotons,'' {\em Physical review letters}, vol.~93, no.~23, p.~230503, 2004.

\bibitem{zhang2024entanglement}
Z.~Zhang, C.~You, O.~S. Maga{\~n}a-Loaiza, R.~Fickler, R.~d.~J. Le{\'o}n-Montiel, J.~P. Torres, T.~S. Humble, S.~Liu, Y.~Xia, and Q.~Zhuang, ``Entanglement-based quantum information technology: a tutorial,'' {\em Advances in Optics and Photonics}, vol.~16, no.~1, pp.~60--162, 2024.

\bibitem{burlakov1999polarization}
A.~Burlakov, M.~Chekhova, O.~Karabutova, D.~Klyshko, and S.~Kulik, ``Polarization state of a biphoton: Quantum ternary logic,'' {\em Physical Review A}, vol.~60, no.~6, p.~R4209, 1999.

\bibitem{fedorov2008spontaneous}
M.~Fedorov, M.~Efremov, P.~Volkov, E.~Moreva, S.~Straupe, and S.~Kulik, ``Spontaneous parametric down-conversion: Anisotropical and anomalously strong narrowing of biphoton momentum correlation distributions,'' {\em Physical Review A}, vol.~77, no.~3, p.~032336, 2008.

\bibitem{fedorov2011entanglement}
M.~Fedorov, P.~Volkov, J.~M. Mikhailova, S.~Straupe, and S.~Kulik, ``Entanglement of biphoton states: qutrits and ququarts,'' {\em New Journal of Physics}, vol.~13, no.~8, p.~083004, 2011.

\bibitem{hurvitz2023frequency}
I.~Hurvitz, A.~Karnieli, and A.~Arie, ``Frequency-domain engineering of bright squeezed vacuum for continuous-variable quantum information,'' {\em Optics Express}, vol.~31, no.~12, pp.~20387--20397, 2023.

\bibitem{shukhin2024two}
A.~Shukhin, I.~Hurvitz, S.~Trajtenberg-Mills, A.~Arie, and H.~Eisenberg, ``Two-dimensional control of a biphoton joint spectrum,'' {\em Optics Express}, vol.~32, no.~6, pp.~10158--10174, 2024.

\bibitem{dosseva2016shaping}
A.~Dosseva, {\L}.~Cincio, and A.~M. Bra{\'n}czyk, ``Shaping the joint spectrum of down-converted photons through optimized custom poling,'' {\em Physical Review A}, vol.~93, no.~1, p.~013801, 2016.

\bibitem{meyer2017limits}
E.~Meyer-Scott, N.~Montaut, J.~Tiedau, L.~Sansoni, H.~Herrmann, T.~J. Bartley, and C.~Silberhorn, ``Limits on the heralding efficiencies and spectral purities of spectrally filtered single photons from photon-pair sources,'' {\em Physical Review A}, vol.~95, no.~6, p.~061803, 2017.

\bibitem{graffitti2018design}
F.~Graffitti, J.~Kelly-Massicotte, A.~Fedrizzi, and A.~M. Bra{\'n}czyk, ``Design considerations for high-purity heralded single-photon sources,'' {\em Physical Review A}, vol.~98, no.~5, p.~053811, 2018.

\bibitem{graffitti2020direct}
F.~Graffitti, P.~Barrow, A.~Pickston, A.~M. Bra{\'n}czyk, and A.~Fedrizzi, ``Direct generation of tailored pulse-mode entanglement,'' {\em Physical Review Letters}, vol.~124, no.~5, p.~053603, 2020.

\bibitem{morrison2022frequency}
C.~L. Morrison, F.~Graffitti, P.~Barrow, A.~Pickston, J.~Ho, and A.~Fedrizzi, ``Frequency-bin entanglement from domain-engineered down-conversion,'' {\em APL Photonics}, vol.~7, no.~6, 2022.

\bibitem{fedorov2014schmidt}
M.~Fedorov and N.~Miklin, ``Schmidt modes and entanglement,'' {\em Contemporary Physics}, vol.~55, no.~2, pp.~94--109, 2014.

\bibitem{sharapova2015schmidt}
P.~Sharapova, A.~M. P{\'e}rez, O.~V. Tikhonova, and M.~V. Chekhova, ``Schmidt modes in the angular spectrum of bright squeezed vacuum,'' {\em Physical Review A}, vol.~91, no.~4, p.~043816, 2015.

\bibitem{just2013transverse}
F.~Just, A.~Cavanna, M.~V. Chekhova, and G.~Leuchs, ``Transverse entanglement of biphotons,'' {\em New Journal of Physics}, vol.~15, no.~8, p.~083015, 2013.

\bibitem{amooei2024efficient}
M.~Amooei, G.~Kulkarni, J.~Upham, and R.~W. Boyd, ``Efficient characterization of spatial {S}chmidt modes of multiphoton entangled states produced from high-gain parametric down-conversion,'' {\em arXiv preprint arXiv:2410.04505}, 2024.

\bibitem{bhattacharjee2022measurement}
A.~Bhattacharjee, N.~Meher, and A.~K. Jha, ``Measurement of two-photon position--momentum {E}instein--{P}odolsky--{R}osen correlations through single-photon intensity measurements,'' {\em New Journal of Physics}, vol.~24, no.~5, p.~053033, 2022.

\bibitem{averchenko2020reconstructing}
V.~A. Averchenko, G.~Frascella, M.~Kalash, A.~Cavanna, and M.~V. Chekhova, ``Reconstructing two-dimensional spatial modes for classical and quantum light,'' {\em Physical Review A}, vol.~102, no.~5, p.~053725, 2020.

\bibitem{kulkarni2018angular}
G.~Kulkarni, L.~Taneja, S.~Aarav, and A.~K. Jha, ``Angular {S}chmidt spectrum of entangled photons: derivation of an exact formula and experimental characterization for noncollinear phase matching,'' {\em Physical Review A}, vol.~97, no.~6, p.~063846, 2018.

\bibitem{kulkarni2022classical}
G.~Kulkarni, J.~Rioux, B.~Braverman, M.~V. Chekhova, and R.~W. Boyd, ``Classical model of spontaneous parametric down-conversion,'' {\em Physical Review Research}, vol.~4, no.~3, p.~033098, 2022.

\bibitem{dixon2012quantum}
P.~B. Dixon, G.~A. Howland, J.~Schneeloch, and J.~C. Howell, ``Quantum mutual information capacity for high-dimensional entangled states,'' {\em Physical review letters}, vol.~108, no.~14, p.~143603, 2012.

\bibitem{luo2019quantum}
Y.-H. Luo, H.-S. Zhong, M.~Erhard, X.-L. Wang, L.-C. Peng, M.~Krenn, X.~Jiang, L.~Li, N.-L. Liu, C.-Y. Lu, {\em et~al.}, ``Quantum teleportation in high dimensions,'' {\em Physical review letters}, vol.~123, no.~7, p.~070505, 2019.

\bibitem{sit2017high}
A.~Sit, F.~Bouchard, R.~Fickler, J.~Gagnon-Bischoff, H.~Larocque, K.~Heshami, D.~Elser, C.~Peuntinger, K.~G{\"u}nthner, B.~Heim, {\em et~al.}, ``High-dimensional intracity quantum cryptography with structured photons,'' {\em Optica}, vol.~4, no.~9, pp.~1006--1010, 2017.

\bibitem{asban2019quantum}
S.~Asban, K.~E. Dorfman, and S.~Mukamel, ``Quantum phase-sensitive diffraction and imaging using entangled photons,'' {\em Proceedings of the National Academy of Sciences}, vol.~116, no.~24, pp.~11673--11678, 2019.

\bibitem{fuenzalida2022resolution}
J.~Fuenzalida, A.~Hochrainer, G.~B. Lemos, E.~A. Ortega, R.~Lapkiewicz, M.~Lahiri, and A.~Zeilinger, ``Resolution of quantum imaging with undetected photons,'' {\em Quantum}, vol.~6, p.~646, 2022.

\bibitem{sharapova2018bright}
P.~Sharapova, O.~Tikhonova, S.~Lemieux, R.~Boyd, and M.~Chekhova, ``Bright squeezed vacuum in a nonlinear interferometer: Frequency and temporal {S}chmidt-mode description,'' {\em Physical Review A}, vol.~97, no.~5, p.~053827, 2018.

\bibitem{straupe2011angular}
S.~Straupe, D.~Ivanov, A.~Kalinkin, I.~Bobrov, and S.~Kulik, ``Angular {S}chmidt modes in spontaneous parametric down-conversion,'' {\em Physical Review A}, vol.~83, no.~6, p.~060302, 2011.

\bibitem{law2004analysis}
C.~Law and J.~Eberly, ``Analysis and interpretation of high transverse entanglement in optical parametric down conversion,'' {\em Physical review letters}, vol.~92, no.~12, p.~127903, 2004.

\bibitem{jha2011partial}
A.~K. Jha, G.~S. Agarwal, and R.~W. Boyd, ``Partial angular coherence and the angular {S}chmidt spectrum of entangled two-photon fields,'' {\em Physical Review A—Atomic, Molecular, and Optical Physics}, vol.~84, no.~6, p.~063847, 2011.

\bibitem{van2006effect}
M.~Van~Exter, A.~Aiello, S.~Oemrawsingh, G.~Nienhuis, and J.~Woerdman, ``Effect of spatial filtering on the {S}chmidt decomposition of entangled photons,'' {\em Physical Review A—Atomic, Molecular, and Optical Physics}, vol.~74, no.~1, p.~012309, 2006.

\bibitem{miatto2012spatial}
F.~M. Miatto, H.~di~Lorenzo~Pires, S.~M. Barnett, and M.~P. van Exter, ``Spatial {S}chmidt modes generated in parametric down-conversion,'' {\em The European Physical Journal D}, vol.~66, pp.~1--11, 2012.

\bibitem{bogaerts_programmable_2020}
W.~Bogaerts, D.~Pérez, J.~Capmany, D.~A.~B. Miller, J.~Poon, D.~Englund, F.~Morichetti, and A.~Melloni, ``Programmable photonic circuits,'' {\em Nature}, vol.~586, no.~7828, pp.~207--216, 2020.

\bibitem{harris2018linear}
N.~C. Harris, J.~Carolan, D.~Bunandar, M.~Prabhu, M.~Hochberg, T.~Baehr-Jones, M.~L. Fanto, A.~M. Smith, C.~C. Tison, P.~M. Alsing, {\em et~al.}, ``Linear programmable nanophotonic processors,'' {\em Optica}, vol.~5, no.~12, pp.~1623--1631, 2018.

\bibitem{sibson2017chip}
P.~Sibson, C.~Erven, M.~Godfrey, S.~Miki, T.~Yamashita, M.~Fujiwara, M.~Sasaki, H.~Terai, M.~G. Tanner, C.~M. Natarajan, {\em et~al.}, ``Chip-based quantum key distribution,'' {\em Nature communications}, vol.~8, no.~1, p.~13984, 2017.

\bibitem{peruzzo2014variational}
A.~Peruzzo, J.~McClean, P.~Shadbolt, M.-H. Yung, X.-Q. Zhou, P.~J. Love, A.~Aspuru-Guzik, and J.~L. O’brien, ``A variational eigenvalue solver on a photonic quantum processor,'' {\em Nature communications}, vol.~5, no.~1, p.~4213, 2014.

\bibitem{carolan2020variational}
J.~Carolan, M.~Mohseni, J.~P. Olson, M.~Prabhu, C.~Chen, D.~Bunandar, M.~Y. Niu, N.~C. Harris, F.~N. Wong, M.~Hochberg, {\em et~al.}, ``Variational quantum unsampling on a quantum photonic processor,'' {\em Nature Physics}, vol.~16, no.~3, pp.~322--327, 2020.

\bibitem{matthews2009manipulation}
J.~C. Matthews, A.~Politi, A.~Stefanov, and J.~L. O'brien, ``Manipulation of multiphoton entanglement in waveguide quantum circuits,'' {\em Nature Photonics}, vol.~3, no.~6, pp.~346--350, 2009.

\bibitem{wang2018multidimensional}
J.~Wang, S.~Paesani, Y.~Ding, R.~Santagati, P.~Skrzypczyk, A.~Salavrakos, J.~Tura, R.~Augusiak, L.~Man{\v{c}}inska, D.~Bacco, {\em et~al.}, ``Multidimensional quantum entanglement with large-scale integrated optics,'' {\em Science}, vol.~360, no.~6386, pp.~285--291, 2018.

\bibitem{harris2017quantum}
N.~C. Harris, G.~R. Steinbrecher, M.~Prabhu, Y.~Lahini, J.~Mower, D.~Bunandar, C.~Chen, F.~N. Wong, T.~Baehr-Jones, M.~Hochberg, {\em et~al.}, ``Quantum transport simulations in a programmable nanophotonic processor,'' {\em Nature Photonics}, vol.~11, no.~7, pp.~447--452, 2017.

\bibitem{miller2013self}
D.~A.~B. Miller, ``Self-configuring universal linear optical component,'' {\em Photonics Research}, vol.~1, no.~1, pp.~1--15, 2013.

\bibitem{miller2013self2}
D.~A.~B. Miller, ``Self-aligning universal beam coupler,'' {\em Optics express}, vol.~21, no.~5, pp.~6360--6370, 2013.

\bibitem{miller2019waves}
D.~A. Miller, ``Waves, modes, communications, and optics: a tutorial,'' {\em Advances in Optics and Photonics}, vol.~11, no.~3, pp.~679--825, 2019.

\bibitem{seyedinnavadeh_determining_2023}
S.~SeyedinNavadeh, M.~Milanizadeh, F.~Zanetto, G.~Ferrari, M.~Sampietro, M.~Sorel, D.~A.~B. Miller, A.~Melloni, and F.~Morichetti, ``Determining the optimal communication channels of arbitrary optical systems using integrated photonic processors,'' {\em Nature Photonics}, pp.~1--7, 2023.

\bibitem{MillerAnalyze2020}
D.~A.~B. Miller, ``Analyzing and generating multimode optical fields using self-configuring networks,'' {\em Optica}, vol.~7, pp.~794--801, 2020.

\bibitem{milanizadeh2021coherent}
M.~Milanizadeh, F.~Toso, G.~Ferrari, T.~Jonuzi, D.~A.~B. Miller, A.~Melloni, and F.~Morichetti, ``Coherent self-control of free-space optical beams with integrated silicon photonic meshes,'' {\em Photonics Research}, vol.~9, no.~11, pp.~2196--2204, 2021.

\bibitem{pai2023experimentally}
S.~Pai, Z.~Sun, T.~W. Hughes, T.~Park, B.~Bartlett, I.~A. Williamson, M.~Minkov, M.~Milanizadeh, N.~Abebe, F.~Morichetti, {\em et~al.}, ``Experimentally realized in situ backpropagation for deep learning in photonic neural networks,'' {\em Science}, vol.~380, no.~6643, pp.~398--404, 2023.

\bibitem{miller2015perfect}
D.~A.~B. Miller, ``Perfect optics with imperfect components,'' {\em Optica}, vol.~2, no.~8, pp.~747--750, 2015.

\bibitem{miller2013establishing}
D.~A.~B. Miller, ``Establishing optimal wave communication channels automatically,'' {\em Journal of Lightwave Technology}, vol.~31, no.~24, pp.~3987--3994, 2013.

\bibitem{roques2024measuring}
C.~Roques-Carmes, S.~Fan, and D.~A. Miller, ``Measuring, processing, and generating partially coherent light with self-configuring optics,'' {\em Light: Science \& Applications}, vol.~13, no.~1, p.~260, 2024.

\bibitem{lu2023frequency}
H.-H. Lu, M.~Liscidini, A.~L. Gaeta, A.~M. Weiner, and J.~M. Lukens, ``Frequency-bin photonic quantum information,'' {\em Optica}, vol.~10, no.~12, pp.~1655--1671, 2023.

\bibitem{nielsen2010quantum}
M.~A. Nielsen and I.~L. Chuang, {\em Quantum computation and quantum information}.
\newblock Cambridge university press, 2010.

\bibitem{hughes2018training}
T.~W. Hughes, M.~Minkov, Y.~Shi, and S.~Fan, ``Training of photonic neural networks through in situ backpropagation and gradient measurement,'' {\em Optica}, vol.~5, no.~7, pp.~864--871, 2018.

\bibitem{kingma2014adam}
D.~P. Kingma and J.~Ba, ``Adam: A method for stochastic optimization.'' arXiv preprint arXiv:1412.6980, 2014.

\bibitem{du2017gradient}
S.~S. Du, C.~Jin, J.~D. Lee, M.~I. Jordan, A.~Singh, and B.~Poczos, ``Gradient descent can take exponential time to escape saddle points,'' {\em Advances in neural information processing systems}, vol.~30, 2017.

\bibitem{lib2022quantum}
O.~Lib and Y.~Bromberg, ``Quantum light in complex media and its applications,'' {\em Nature Physics}, vol.~18, no.~9, pp.~986--993, 2022.

\bibitem{burgwal2017using}
R.~Burgwal, W.~R. Clements, D.~H. Smith, J.~C. Gates, W.~S. Kolthammer, J.~J. Renema, and I.~A. Walmsley, ``Using an imperfect photonic network to implement random unitaries,'' {\em Optics Express}, vol.~25, no.~23, pp.~28236--28245, 2017.

\bibitem{mair2001entanglement}
A.~Mair, A.~Vaziri, G.~Weihs, and A.~Zeilinger, ``Entanglement of the orbital angular momentum states of photons,'' {\em Nature}, vol.~412, no.~6844, pp.~313--316, 2001.

\bibitem{erhard2018twisted}
M.~Erhard, R.~Fickler, M.~Krenn, and A.~Zeilinger, ``Twisted photons: new quantum perspectives in high dimensions,'' {\em Light: Science \& Applications}, vol.~7, no.~3, pp.~17146--17146, 2018.

\bibitem{kovlakov2018quantum}
E.~Kovlakov, S.~Straupe, and S.~Kulik, ``Quantum state engineering with twisted photons via adaptive shaping of the pump beam,'' {\em Physical Review A}, vol.~98, no.~6, p.~060301, 2018.

\bibitem{hu2021chip}
Y.~Hu, M.~Yu, D.~Zhu, N.~Sinclair, A.~Shams-Ansari, L.~Shao, J.~Holzgrafe, E.~Puma, M.~Zhang, and M.~Lon{\v{c}}ar, ``On-chip electro-optic frequency shifters and beam splitters,'' {\em Nature}, vol.~599, no.~7886, pp.~587--593, 2021.

\bibitem{zhu2022spectral}
D.~Zhu, C.~Chen, M.~Yu, L.~Shao, Y.~Hu, C.~Xin, M.~Yeh, S.~Ghosh, L.~He, C.~Reimer, {\em et~al.}, ``Spectral control of nonclassical light pulses using an integrated thin-film lithium niobate modulator,'' {\em Light: Science \& Applications}, vol.~11, no.~1, p.~327, 2022.

\bibitem{lingaraju2022bell}
N.~B. Lingaraju, H.-H. Lu, D.~E. Leaird, S.~Estrella, J.~M. Lukens, and A.~M. Weiner, ``Bell state analyzer for spectrally distinct photons,'' {\em Optica}, vol.~9, no.~3, pp.~280--283, 2022.

\bibitem{werner1995ultrashort}
M.~Werner, M.~Raymer, M.~Beck, and P.~Drummond, ``Ultrashort pulsed squeezing by optical parametric amplification,'' {\em Physical Review A}, vol.~52, no.~5, p.~4202, 1995.

\bibitem{arzani2018versatile}
F.~Arzani, C.~Fabre, and N.~Treps, ``Versatile engineering of multimode squeezed states by optimizing the pump spectral profile in spontaneous parametric down-conversion,'' {\em Physical Review A}, vol.~97, no.~3, p.~033808, 2018.

\bibitem{u2006generation}
A.~B. U'Ren, C.~Silberhorn, R.~Erdmann, K.~Banaszek, W.~P. Grice, I.~A. Walmsley, and M.~G. Raymer, ``Generation of pure-state single-photon wavepackets by conditional preparation based on spontaneous parametric downconversion,'' {\em arXiv preprint quant-ph/0611019}, 2006.

\bibitem{CHRIST2013351}
A.~Christ, A.~Fedrizzi, H.~Hübel, T.~Jennewein, and C.~Silberhorn, ``Chapter 11 - parametric down-conversion,'' in {\em Single-Photon Generation and Detection} (A.~Migdall, S.~V. Polyakov, J.~Fan, and J.~C. Bienfang, eds.), vol.~45 of {\em Experimental Methods in the Physical Sciences}, pp.~351--410, Academic Press, 2013.

\bibitem{presutti2024highly}
F.~Presutti, L.~G. Wright, S.-Y. Ma, T.~Wang, B.~K. Malia, T.~Onodera, and P.~L. McMahon, ``Highly multimode visible squeezed light with programmable spectral correlations through broadband up-conversion,'' {\em arXiv preprint arXiv:2401.06119}, 2024.

\bibitem{boucher2021engineering}
P.~Boucher, H.~Defienne, and S.~Gigan, ``Engineering spatial correlations of entangled photon pairs by pump beam shaping,'' {\em Optics Letters}, vol.~46, no.~17, pp.~4200--4203, 2021.

\bibitem{luo2017chip}
R.~Luo, H.~Jiang, S.~Rogers, H.~Liang, Y.~He, and Q.~Lin, ``On-chip second-harmonic generation and broadband parametric down-conversion in a lithium niobate microresonator,'' {\em Optics express}, vol.~25, no.~20, pp.~24531--24539, 2017.

\bibitem{chapman2024chip}
R.~J. Chapman, T.~Kuttner, J.~Kellner, A.~Sabatti, A.~Maeder, G.~Finco, F.~Kaufmann, and R.~Grange, ``On-chip quantum interference between independent lithium niobate-on-insulator photon-pair sources,'' {\em arXiv preprint arXiv:2404.08378}, 2024.

\bibitem{zielnicki2018joint}
K.~Zielnicki, K.~Garay-Palmett, D.~Cruz-Delgado, H.~Cruz-Ramirez, M.~F. O’Boyle, B.~Fang, V.~O. Lorenz, A.~B. U’Ren, and P.~G. Kwiat, ``Joint spectral characterization of photon-pair sources,'' {\em Journal of Modern Optics}, vol.~65, no.~10, pp.~1141--1160, 2018.

\bibitem{defienne2016two}
H.~Defienne, M.~Barbieri, I.~A. Walmsley, B.~J. Smith, and S.~Gigan, ``Two-photon quantum walk in a multimode fiber,'' {\em Science advances}, vol.~2, no.~1, p.~e1501054, 2016.

\bibitem{defienne2018adaptive}
H.~Defienne, M.~Reichert, and J.~W. Fleischer, ``Adaptive quantum optics with spatially entangled photon pairs,'' {\em Physical review letters}, vol.~121, no.~23, p.~233601, 2018.

\bibitem{wolterink2016programmable}
T.~A. Wolterink, R.~Uppu, G.~Ctistis, W.~L. Vos, K.-J. Boller, and P.~W. Pinkse, ``Programmable two-photon quantum interference in 10 3 channels in opaque scattering media,'' {\em Physical Review A}, vol.~93, no.~5, p.~053817, 2016.

\bibitem{cao2020long}
Y.~Cao, Y.-H. Li, K.-X. Yang, Y.-F. Jiang, S.-L. Li, X.-L. Hu, M.~Abulizi, C.-L. Li, W.~Zhang, Q.-C. Sun, {\em et~al.}, ``Long-distance free-space measurement-device-independent quantum key distribution,'' {\em Physical Review Letters}, vol.~125, no.~26, p.~260503, 2020.

\bibitem{lib2020real}
O.~Lib, G.~Hasson, and Y.~Bromberg, ``Real-time shaping of entangled photons by classical control and feedback,'' {\em Science advances}, vol.~6, no.~37, p.~eabb6298, 2020.

\bibitem{shekel2024shaping}
R.~Shekel, O.~Lib, and Y.~Bromberg, ``Shaping entangled photons through arbitrary scattering media using an advanced wave beacon,'' {\em Optica Quantum}, vol.~2, no.~5, pp.~303--309, 2024.

\bibitem{bogdanov2006schmidt}
A.~Y. Bogdanov, Y.~I. Bogdanov, and K.~Valiev, ``Schmidt modes and entanglement in continuous-variable quantum systems,'' {\em Russian Microelectronics}, vol.~35, pp.~7--20, 2006.

\bibitem{bravo2020quantum}
C.~Bravo-Prieto, D.~Garc{\'\i}a-Mart{\'\i}n, and J.~I. Latorre, ``Quantum singular value decomposer,'' {\em Physical Review A}, vol.~101, no.~6, p.~062310, 2020.

\bibitem{wang2021variational}
X.~Wang, Z.~Song, and Y.~Wang, ``Variational quantum singular value decomposition,'' {\em Quantum}, vol.~5, p.~483, 2021.

\bibitem{fickler2014interface}
R.~Fickler, R.~Lapkiewicz, M.~Huber, M.~P. Lavery, M.~J. Padgett, and A.~Zeilinger, ``Interface between path and orbital angular momentum entanglement for high-dimensional photonic quantum information,'' {\em Nature communications}, vol.~5, no.~1, p.~4502, 2014.

\bibitem{lightman2017miniature}
S.~Lightman, G.~Hurvitz, R.~Gvishi, and A.~Arie, ``Miniature wide-spectrum mode sorter for vortex beams produced by 3d laser printing,'' {\em Optica}, vol.~4, no.~6, pp.~605--610, 2017.

\end{thebibliography}

\end{document}